\documentclass{article}
\usepackage[affil-it]{authblk}
\usepackage{amsmath}
\usepackage{amsfonts}

\newcommand{\PB}[2]{\mbox{$\left\{#1,#2\right\}$}}

\newcommand{\beq}{\begin{eqnarray}}
\newcommand{\eeq}{\end{eqnarray}}
\newcommand{\nn}{\nonumber}

\newcommand{\Tr}{{\mathrm{Tr}}}

\newcommand{\vsp}[1]{\mbox{$\mathbf{#1}$}}

\title{Causal Poisson bracket via deformation quantization}

\author[1]{Jasel Berra-Montiel\footnote{\texttt{jberra@fc.uaslp.mx}}}

\affil[1]{Facultad de Ciencias, Universidad Aut\'onoma de San Luis 
Potos\'{\i}
Av. Salvador Nava S/N Zona Universitaria, CP 78290, San 
Luis Potos\'{\i}, SLP, M\'exico}

\author[1,2]{Alberto Molgado\footnote{\texttt{molgado@fc.uaslp.mx}}}
\affil[2]{Dual CP Institute of High Energy Physics, M\'exico  
}

\author[3]{C\'esar D.~Palacios-Garc\'ia\footnote{\texttt{cpalacios@dec1.ifisica.uaslp.mx}}}
\affil[3]{Instituto de F\'isica ``Manuel Sandoval Vallarta", Universidad 
Aut\'onoma de San Luis Potos\'i  
\'Alvaro Obreg\'on 64, 78000 San Luis Potos\'i,
SLP, M\'exico   
}

\begin{document}


\markboth{J.~Berra-Montiel, A.~Molgado and C.~D.~Palacios-Garc\'ia}{Causal Poisson bracket via deformation quantization}

%
%

\maketitle


\begin{abstract}
Starting with the 
well-defined product of quantum fields at two spacetime points, we explore
an associated Poisson structure for classical field theories within the deformation 
quantization formalism.  We realize that the induced star-product is naturally related 
to the standard Moyal product through an 
appropriate causal Green's functions connecting points in the 
space of classical solutions to the equations of motion.  Our results resemble the 
Peierls-DeWitt bracket analyzed in the multisymplectic context.  
Once our star-product is defined we are able to apply the Wigner-Weyl map in order to 
introduce a generalized 
version of Wick's theorem.  Finally, we include some 
examples to explicitly test our method:  the 
real scalar field, the bosonic string and a
physically motivated
nonlinear particle model.
For the field theoretic models we have encountered 
causal generalizations
of the creation/annihilation relations, and 
also a causal generalization of the Virasoro algebra
for the bosonic string.
For the nonlinear particle case,
we use the approximate solution in terms of the 
Green's function in order to construct a well-behaved
 causal bracket. 
\end{abstract}


\section{Introduction}
\label{sec:intro}

Standard quantization procedures for field 
theories rely to some extent  
on a Poisson structure at the classical 
level.  
Even though a classical field theory 
may be completely understood at either the 
Lagrangian or the Hamiltonian formalisms and, in spite of the mathematical elegance of both approaches, a covariant 
Poisson formulation for classical 
fields has not been completely embraced by 
a vast community of physicists.  
As it may be suspected, this is at the very heart of most of the relevant physical 
systems, including all of the fundamental interactions within the standard model, 
gravitation and string theory, to mention 
some.
In this way, one naturally starts by considering 
a covariant classical field theory for which 
one may apply certain standard rules in order to 
get a quantized version.  However, this rules 
impose at some point a preferred foliation of spacetime
in order to fulfill the quantization programme,
thus apparently hiding the covariant character of 
a given field theory. 

In this direction, the deformation quantization approach was introduced in \cite{BFFLS} as an alternative procedure for standard quantization. 
The deformation quantization programme has shown to be a mathematical consistent
tool for the understanding of quantum systems ranging form standard 
quantum mechanics to quantum aspects of general Lie algebraic structures.
In this formalism, quantizing a classical system
simply consists on a deformation of the corresponding algebraic structures such as the algebra of smooth functions defined on the classical phase space. For details, we refer 
the reader to the seminal papers \cite{BFFLS,FlatoView}, and the 
reviews~\cite[and references therein]{Dito,danielstern,QMOPS,Zachos,
Curtright} for a wide range of applications and recent developments.   

Our major objective in this paper is to develop, within the deformation quantization formalism, a legitimate 
algebraic causal Poisson
bracket for classical field theories.  
On the one hand, for linear theories, we adopt the 
familiar Kirchhoff representation~\cite{Living,
Hadamard,Baer,Sciama,Waldmann} which states that given a 
field and its normal derivative at a 
given hypersurface we may find the value of 
the field at any 
causally connected point in the chronological  future of the original point.  This 
result is based on the construction of 
an appropriate Green's function and, 
in principle, holds even for curved spacetimes. 
On the other hand, even though the Kirchhoff representation
is not valid for nonlinear theories, we may introduce 
a Green's function which gives us the approximate solution in a suitable region 
for which the causal Poisson bracket becomes 
meaningful~\cite{Frasca1,Frasca2,Epperson,Weike,Taigbenu}. 
Thus our claim is that 
to the unambiguous 
well-defined product of two 
quantum field
operators evaluated at different but causally connected spacetime
points it is possible to assign a 
correspondent 
classical causal Poisson structure.   
Certainly, the Wigner function 
allows us to map both quantum field operators to 
operators evaluated at different points belonging to the 
same hypersurface by means of the Stratonovich-Weyl quantizer which admits not only continuous differentiable functions but also distributions.
Further, the so-called correspondence principle indicates that 
the resulting star-product is interrelated to a well-defined 
classical causal Poisson 
bracket at two different spacetime points given in terms of the causal  
Green's function associated to the field 
equations for both linear and nonlinear systems.  
The introduced bracket also reduces to 
the standard Poisson bracket whenever we
consider the two spacetime points in the 
same spatial hypersurface, that is, in the equal-time limit of field theory.

This causal Poisson 
bracket results equivalent to the 
covariant
Peierls-DeWitt bracket~\cite{Peierls,DeWitt,Dewitt1} as far 
as linear field
theories are considered (see also~\cite{Khavkine} for an excellent 
review on this topic).  Examples of 
these linear field theories are given 
by non-interacting theories, harmonic Lagrangians and self-adjoint functionals, examples which encompass a huge amount of physically
interesting field theories~\cite{Witten,Kijowski,Kanatchikov,Helein}.  
Nonetheless, whenever we 
consider nonlinear field theories our bracket 
diverges from the Peierls-DeWitt bracket, 
as the causal Poisson
bracket introduced is only related to the first variation of the action, 
and thus do not depend on a linearized version of the 
field equations. 
In this way, the difference among the brackets 
may be clearer if we 
bear in mind that for the Peierls bracket the involved 
causal Green's functions
turn out to be Jacobi fields, while for the causal bracket the causal Green's functions
are not necessarily Jacobi fields.
In this sense, for nonlinear field theories,
the causal bracket is not compulsorily covariant but it 
preserves the causal structure
in an approximate manner in a suitable region 
of interest. 
Similar causal Poisson bracket 
structures has been
implemented in a variety of contexts, 
including the conformal field theoretical
WZNW model, the causal algebras, 
the localization of particles in QFT, to mention 
some~\cite{FJW,FJ,JW,Westra,Koksma}.

In the case of field theories with interactions,
a perturbative approximation must be 
considered, as in standard quantum field theory.
However, our developed star-product lead us 
to obtain a generalization of Wick's theorem
for the product of field operators at different 
spacetime points.  This generalization 
involves convenient contractions of the field operators 
with the causal Green's functions involved.  
Besides, we are able, by means 
of an isomorphism between star-products, to introduce 
a relation between our causal Green's function and 
Feynman's propagator, thus interpolating  
both 
approaches.  These results resemble analogous developments 
found in deformation quantization from an 
algebraic quantum field theory 
perspective~\cite{Freden,Freden2,
Hirshfeld1,Hirshfeld2}.

Finally, we test the 
causal Poisson bracket formalism
for some examples.  Firstly,
we analyze in detail the real scalar field.
In particular, we find that the classical 
Poisson brackets may be extended to allow
relations among the annihilation and creation coefficients at different 
spacetime hypersurfaces, generalizing the 
conventional relations at the equal-time
limit analyzed in canonical quantization.
Secondly, we also investigate the bosonic 
string.  In this case, we also find a causal
version of the Poisson brackets for the 
mode expansion coefficients, which in turn 
lead us to a generalized version of the 
Virasoro algebra at two different spatial 
hypersurfaces.  Lastly, we also analyze a 
nonlinear example for which a causal 
Green's function may be introduced appropriately.

The paper is organized as follows.
In Section~\ref{sec:review} we give a 
brief review of deformation quantization
in order to set the notational conventions
used in the subsequent sections.  In Section~\ref{sec:CovPoissonStructure} 
we introduce the causal Poisson 
structure for field theory, and study its 
relevant properties.  We test the 
causal Poisson bracket by developing in
detail some specific examples in 
Section~\ref{sec:Examples}.  We include 
some concluding remarks in 
Section~\ref{sec:conclu}.  Finally, 
we leave technical demonstrations of some mathematical properties of the causal
bracket to~\ref{sec:MathProps}.

\section{Deformation quantization for field theory}
\label{sec:review}

In classical mechanics, the phase space is given by a Poisson manifold $\mathcal{M}$, 
together with an antisymmetric Poisson tensor $\alpha^{ij}$, which endows the commutative algebra of complex-valued smooth functions $C^{\infty}(\mathcal{M})$ with a Lie algebraic structure by means of the bracket $\PB{\cdot}{\cdot}:C^{\infty}(\mathcal{M})\times C^{\infty}(\mathcal{M})\rightarrow C^{\infty}(\mathcal{M})$ explicitly given by
\begin{equation}
\left\lbrace f,g\right\rbrace = \alpha^{ij}\partial_{i}f\partial_{j}g, 
\end{equation}
which, besides skew-symmetry and bilinearity, satisfies the Jacobi identity
\beq
\label{eq:Jacobi}
\PB{f}{\PB{g}{h}}=\PB{\PB{f}{g}}{h} + \PB{g}{\PB{f}{h}} \,,
\eeq
and the compatibility Leibnizian condition 
\begin{equation}
\label{eq:Leibniz}
\left\lbrace f,gh\right\rbrace =\left\lbrace f,g\right\rbrace h+g\left\lbrace f,h \right\rbrace \,,  
\end{equation}
showing that the Poisson bracket is a derivation under both, the Poisson bracket itself and the 
standard commutative product of functions. Whenever the 
Poisson tensor $\alpha^{ij}$ is 
non-degenerate, the manifold $\mathcal{M}$ is said to be a 
symplectic manifold.  Non-degenerate Poisson tensors 
mainly comprises systems 
without local symmetries, although for gauge invariant systems a 
symplectic manifold may be constructed in the so-called reduced phase space.  
For symplectic manifolds, the Jacobi 
identity turns out to have an immediate geometrical meaning, since it is equivalent 
to the closedness of the $2$-form
\begin{equation}
\omega=\frac{1}{2}\omega_{ij}dx^{i}\wedge dx^{j},
\end{equation}
\noindent where $\omega_{ij}$ denotes the inverse matrix to $\alpha^{ij}$.  

A deformation quantization stands for an associative algebraic 
structure $\mathcal{A}:=(A(\mathcal{M}),\star)$ on the space $A(\mathcal{M}):=C^{\infty}(\mathcal{M})[[\hbar]]$ of formal power series in a formal parameter $\hbar$ with respect to an associative product, the so-called star-product $\star$, satisfying for each 
$f,\ g \in C^\infty(\mathcal{M})$ the following properties
\begin{enumerate}
\item Locality property:
\beq
\nn
f\star g=\sum_{k=0}^{\infty}\left( \frac{i\hbar}{2}\right)^{k}C_{k}(f,g)  \,,
\eeq
where $C_{k}(f,g)$ are a sequence of bidifferential operators.
\item Deformation property:  The star-product is a formal associative deformation of the classical commutative product, that, is
\beq
\nn
C_{0}(f,g)=fg \,.
\eeq
\item Correspondence principle:  The star-commutator allows us to define a formal deformation of the Poisson bracket 
\beq
\nn
C_{1}(f,g)-C_{1}(g,f)=i\hbar\left\lbrace f,g\right\rbrace. 
\eeq
\end{enumerate}  

Besides, two star-products, $\star$ and 
$\star'$, are said to be equivalent if there is an isomorphism between the algebras $\mathcal{A}=(A,\star)$ and 
$\mathcal{A}'=(A,\star')$ given by a formal differential operator $T=\mathbb{I}+\sum_{r}\hbar^{r}T_{r}$, where $\mathbb{I}$ stands for the identity operator and each $T_{r}$ is a differential operator which is null on constants, and such that the differential
operator $T$ follows
\begin{equation}
\label{eq:StarEquiv}
T(f\star' g)=(Tf)\star (Tg) \,.
\end{equation}  
This equivalence is related to the operator ordering ambiguity 
in ordinary quantum mechanics.  In this way, defining 
a new star-product may be interpreted as a change in the ordering prescription in a quantum theory.  Besides, this equivalence
will be also relevant to understand the connection 
between our causal propagator and the Feynman propagator in 
standard quantum field theory, as we will see below. 

In deformation quantization, the algebra of quantum 
observables turns out to be particularly simple, since it 
is made up by the set of real-valued functions on the phase space. In this manner there is no need of a Hilbert space as in the traditional operator approach, hence avoiding the more difficult problem about domains of unbounded operators. 
Furthermore, a 
very important point in deformation quantization comes
from the existence of the Kontsevich theorem 
which provides a universal procedure to construct 
a well defined star-product starting with an arbitrary classical system, as it states that an arbitrary Poisson manifold admits a deformation quantization~\cite{Kontsevich,conjetura,fedosov,alfonso}.  


Hereinafter, let us specialize our considerations so far to the case of an arbitrary field theory on four-dimensional Minkowski 
spacetime $\mathcal{M}$.  We will follow as close as possible the notation in 
references~\cite{Dito1,Compean}. 
As customarily, we will denote the canonical variables  as 
$\Phi^{I}(x)$ and $\Pi_{I}(x)$, where the index $I$
stands for the set of internal indices, and 
depends on the nature of 
each field (and may be omitted 
when possible), and spacetime points $x=(x^{0},x^{i})\in 
\mathcal{M}$,  $i=1,2,3$, standing for spatial indices. 
In deformation quantization, a common 
starting point will be to define either the Weyl map, 
or its inverse, 
the quasi-probabilistic Wigner function, both setting a relation 
between classical observables and quantum 
operators~\cite{Nachbagauer,Curtright1}. We will thus start by constructing the Weyl map.  
Let $F\left[\Phi,\Pi \right]$ be an arbitrary 
functional defined on the phase space $\Gamma(\mathcal{M})$ associated to 
$\mathcal{M}$.  We define its Fourier transformation by
\begin{equation}
\tilde{F}[\lambda,\mu]=\int \mathcal{D}\Phi\mathcal{D} \Pi\exp\left\lbrace -i\int dx\left( \lambda(x)\cdot \Phi(x)+\mu(x)\cdot\Pi(x) \right) \right\rbrace F[\Phi,\Pi]  \,, 
\end{equation}
where the formal functional measures are given by 
$\mathcal{D}\Phi=\prod_{x}d\Phi(x)$, and $\mathcal{D}\Pi=\prod_{x}d\Pi(x)$, respectively, and the central dot stands for contraction on the 
appropriate indices. Thus, the Weyl map in phase space is given by the quantum operator $\hat{F}$ associated to $F[\Phi,\Pi]$
\beq
\label{eq:WeylMap1}
\hat{F}
:=  
W\left( F[\Phi,\Pi]\right) 
=  
\int \mathcal{D}\left( \frac{\lambda}{2\pi}\right)\mathcal{D}\left( \frac{\mu}{2\pi}\right)\tilde{F}[\lambda,\mu]\hat{U}[\lambda,\mu]   \,,   
\eeq
where $\hat{U}[\lambda,\mu]$ stands for the unitary operator
\beq
\label{eq:UnitaryU}
\hat{U}[\lambda,\mu]=\exp\left\lbrace i \int dx \left( \lambda(x)\cdot\hat{\Phi}(x)+\mu(x)\cdot\hat{\Pi}(x)\right) \right\rbrace \,,
\eeq
\noindent being $\hat{\Phi}$ and $\hat{\Pi}$ field operators  satisfying $\hat{\Phi}(x)\rangle{\Phi(x)}=\Phi(x)\rangle{\Phi(x)}$ 
and $\hat{\Pi}(x)\rangle{\Pi(x)}=\Pi(x)\rangle{\Pi(x)}$, respectively.
As shown in~\cite{Compean}, by employing the completeness relations $\int \mathcal{D}\Phi\rangle{\Phi}\langle{\Phi}=\hat{1}$ and $ \int \mathcal{D}\left( \frac{\Pi}{2\pi\hbar}\right) \rangle{\Pi}\langle{\Pi}=\hat{1}$, it is easy to check that the 
operator~(\ref{eq:UnitaryU}) obeys the two 
very important properties
\beq
\label{eq:TrU1}
\Tr\left(\hat{U}[\lambda,\mu]\right) 
& = &  
\int\mathcal{D}\Phi\langle{\Phi}\hat{U}[\lambda,\mu]\rangle{\Phi}=\delta\left( \frac{\hbar\lambda}{2\pi}\right)\delta\left( \mu\right)\,,   \\
\label{Eq:TrU2}
\Tr\left( \hat{U}^{\dagger}[\lambda,\mu]\hat{U}[\lambda',\mu']\right)  
& = &  
\delta\left(\frac{\hbar}{2\pi}(\lambda-\lambda') \right)\delta(\mu-\mu') \,,\hspace{3ex}
\eeq
where the $\delta$'s stand for Dirac deltas.
Relations~(\ref{eq:TrU1}) and~(\ref{Eq:TrU2}) will be relevant
in order to construct the quasi-probabilistic Wigner function, 
which assigns a  classical observable to a given quantum 
operator.  Before constructing the Wigner function we note that 
the Weyl quantization rule~(\ref{eq:WeylMap1}) may be written as 
\begin{equation}
\label{eq:WeylMap2}
\hat{F}=W\left( F[\Phi,\Pi]\right)=\int\mathcal{D}\Phi \mathcal{D}\left( \frac{\Pi}{2\pi\hbar}\right) F[\Phi,\Pi]\hat{\Omega}[\Phi,\Pi], 
\end{equation}
\noindent where the operator $\hat{\Omega}[\Phi,\Pi]$ denotes 
the standard 
Stratonovich-Weyl quantizer for quantum field theory
\begin{equation}
\label{eq:StratoW}
\hat{\Omega}[\Phi,\Pi]=\int\mathcal{D}\left( \frac{\hbar\lambda}{2\pi}\right)\mathcal{D}\mu\exp\left\lbrace -i\int dx\left( \lambda(x)
\cdot\Phi(x)+\mu(x)\cdot\Pi(x)\right)  \right\rbrace \hat{U}[\lambda,\mu]  . 
\end{equation}
Bearing in mind relations~(\ref{eq:TrU1}) and~(\ref{Eq:TrU2}), 
it is straightforward to check that the Stratonovich-Weyl quantizer $\hat{\Omega}[\Phi,\Pi]$ satisfies the identities
\beq
\label{eq:SW1}
\hat{\Omega}^{\dagger}[\Phi,\Pi] 
& = & 
\hat{\Omega}[\Phi,\Pi] \,,\\
\label{eq:SW2} 
\Tr\left( \hat{\Omega}[\Phi,\Pi]\right)
& = & 
1 \,,  \\
\label{eq:SW3} 
\Tr\left( \hat{\Omega}[\Phi,\Pi]\hat{\Omega}[\Phi',\Pi']\right)
& = & 
\delta(\Phi-\Phi')\delta\left(\frac{\Pi-\Pi'}{2\pi\hbar} \right) \,.  
\eeq
In this notation, the Wigner function simply reads
\begin{equation}
\label{eq:Wigner}
F[\Phi,\Pi] = W^{-1}(\hat{F}) =\Tr\left( \hat{\Omega}[\Phi,\Pi]\hat{F}\right) \,. 
\end{equation}

The next step is to construct a star-product which encloses
an specific ordering prescription, as discussed before.  We will follow here the 
standard Weyl-Moyal ordering~\cite{Hirshfeld2,Dito1}. In order to define the 
field theoretical Moyal 
star-product, let $F_{1}=F_{1}[\Phi,\Pi]$ and $F_{2}=F_{2}[\Phi,\Pi]$ be some functionals on the phase space $\Gamma(\mathcal{M})$, whose corresponding field operators, obtained through the Weyl map~(\ref{eq:WeylMap2}), are $\hat{F}_{1}=W(F_{1})$ and $\hat{F}_{2}=W(F_{2})$, respectively. Thus, 
the Moyal product is defined by means of the convolution relation
\beq
W(F_1\star F_2)=W(F_1)W(F_2) \,,
\eeq
setting 
the functional 
corresponding to the product of two field operators via 
the Wigner function~(\ref{eq:Wigner}) as
\beq
\hspace{-2ex}
(F_{1}\star F_{2})[\Phi,\Pi] & = & W^{-1}(W(F_1)W(F_2))=W^{-1}(\hat{F}_{1}\hat{F_{2}}) =  \Tr\left( \hat{\Omega}[\Phi,\Pi]\hat{F}_{1}\hat{F}_{2}\right), 
\label{eq:MoyalTr}
\eeq
\noindent which may be explicitly written in its integral 
representation as 
\beq
\label{eq:MoyalIntegral}
(F_{1}\star F_{2})[\Phi,\Pi] & = & \int\mathcal{D}\Phi'\mathcal{D}\Phi''\mathcal{D}\left(\frac{ \Pi'}{\pi\hbar}\right)\mathcal{D}\left( \frac{\Pi''}{\pi\hbar}\right)F_{1}[\Phi',\Pi']F_{2}[\Phi'',\Pi''] \nn\\
 & & \times\exp \left\lbrace \frac{2i}{\hbar}\int dx \left( (\Phi-\Phi')\cdot(\Pi-\Pi'')-(\Phi-\Phi'')\cdot(\Pi-\Pi')\right) \right\rbrace \,. \nn\\
\eeq
Finally, using the Taylor series expansion for the 
functionals  $F_{1}$ and $F_{2}$ we obtain the well-known expression
\begin{equation}
\label{eq:MoyalSTD}
(F_{1}\star F_{2})=F_{1}[\Phi,\Pi]\exp\left\lbrace \frac{i\hbar}{2}\stackrel{\leftrightarrow}{\mathcal{P}}\right\rbrace F_{2}[\Phi,\Pi],
\end{equation}
\noindent where $\stackrel{\leftrightarrow}{\mathcal{P}}$ 
stands for the bidirectional functional derivative operator
\begin{equation}
\stackrel{\leftrightarrow}{\mathcal{P}}=\int dx\left( \frac{\overleftarrow{\delta}}{\delta\Phi(x)}\cdot\frac{\overrightarrow{\delta}}{\delta\Pi(x)}- \frac{\overleftarrow{\delta}}{\delta\Phi(x)}\cdot\frac{\overrightarrow{\delta}}{\delta\Pi(x)}\right) \,.  
\end{equation}
It is straightforward to prove that the Moyal 
star-product~(\ref{eq:MoyalSTD}) follows properties (i) to 
(iii) stated before in this section.

\section{Causal Poisson structure for field theory}
\label{sec:CovPoissonStructure}


Our main aim in this section will be to find a 
classical Poisson structure which corresponds to the product of two field operators evaluated at different spacetime points $\hat{\Phi}(x_{1})\hat{\Phi}(x_{2})$, where $x_{1},x_{2}\in \mathcal{M}$. 
To start, let us 
consider a spacelike hypersurface $\Sigma$, and suppose that the values for a field 
$\Phi(x')$ satisfying the Lagrange equations of motion at a point $x'\in\mathcal{M}$, and its normal derivative $n^{\alpha'}\nabla_{\alpha'}\Phi(x')$ are known on $\Sigma$, then the value of the scalar field at a different point 
$x\in\mathcal{M}$ lying on the future of the hypersurface $\Sigma$\footnote{For simplicity, and without 
loss of generality,
we will consider the point $x$ lying in the 
interior of the future light cone of the point
$x'$.  This consideration, however is not 
fundamental, as the causal Green's function considered 
completely
determines the causal structure of the theory.} is given by the formula
\beq
\label{eq:Kirc}
\Phi^{I}(x)=-\int_{\Sigma}\left( \tilde{G}^{IJ}(x,x')\nabla^{\alpha'}\Phi^{J}(x')-\Phi^{J}(x')\nabla^{\alpha'}\tilde{G}^{IJ}(x,x')\right)d\Sigma_{\alpha'}, 
\eeq 
where this formula may be interpreted as exact for linear 
systems and as a well-behaved approximation  for nonlinear systems under 
appropriate boundary 
conditions~\cite{Frasca1,Frasca2,Taigbenu}.  Also, here $\Phi^I(x)$ and $\Phi^J(x')$ stands for the fields evaluated at two 
causally 
connected points
$x,x'\in\mathcal{M}$, respectively,
and $d\Sigma_{\alpha'}$ is the surface element defined on $\Sigma$. It follows that if $n_{\alpha'}$ is the future time-like unit normal, then $d\Sigma_{\alpha'}=n_{\alpha'}dS$, with $dS$ representing the invariant  volume element definded on $\Sigma$. Finally in~(\ref{eq:Kirc}), the causal Green's function $\tilde{G}^{IJ}(x,x')$ is given by 
\beq
\label{eq:CausalGreen}
\tilde{G}^{IJ}(x,x'):=G^{+IJ}(x,x')-G^{-IJ}(x,x')
 \,,
\eeq 
where $G^{+IJ}(x,x')$ and $G^{-IJ}(x,x')$ denote the advanced and retarded Green's function 
associated to the Euler-Lagrange operator, respectively.  
The causal Green's function 
$\tilde{G}^{IJ}(x,x')$ and its complex 
conjugate $(\tilde{G}^{IJ}(x,x'))^*$  
follow the symmetry relations 
\beq
\tilde{G}^{IJ}(x,x') & = &-\tilde{G}^{JI}(x',x) \,,  \\
(\tilde{G}^{IJ})^*(x,x') & = &\tilde{G}^{IJ}(x,x') \,.
\eeq
These 
relations may be checked straightforwardly as a consequence of the reciprocity conditions of the advanced and retarded Green's function
\beq
G^{\pm IJ}(x,x') & = & G^{\mp JI}(x',x) \,, 
\eeq
as discussed in~\cite{Dewitt1}.  
At this point, it is important to mention that 
for a field theory on a flat spacetime 
the causal properties 
of the advanced (retarded) Green's functions 
only has support on the past (future) light cone of a given point $x'$, while the situation 
is subtler for the case of curved spacetimes.
Indeed, for a field theory on a curved spacetime,
its support is extended to consider also 
the interior points of the light cone
due to the fact that in curved spacetime waves propagate at all
speeds equal or smaller than the maximum speed as a result of the 
effect caused by
the interaction between the fields and the curvature of the spacetime.  This has as a consequence a non-regular and 
non-necessarily skew-symmetric causal Green's function.  
Nevertheless, a new two-point function may be added to the causal Green's
function in order to obtain a regular, skew-symmetric causal propagator~\cite{Living,Detwe}.
In this work, we will thus consider flat spacetimes, and in this 
manner we consider the advanced (retarded) Green's function is therefore non-vanishing whenever 
$x\in\mathcal{M}$ belongs to the chronological past (future) of 
$x'\in\mathcal{M}$.   

For convenience, from now on we will adopt numerical subindex notation for causally connected spacetime points, that is, we will denote $x_k\in\Sigma_k,\ k\in\mathbb{N}^+$, where 
$\Sigma_k$ stands for the temporal hypersurface labeled 
by the time $t=t_k$. 
By using the deformation quantization approach
as stated in the last section, the formula~(\ref{eq:Kirc}), and taking the normal derivative $n^{\alpha'}\nabla_{\alpha'}\Phi^{I}(x')$ as a timelike directed derivative 
$\partial \Phi^{I}(x')/\partial x'^{0}$, we obtain through the Weyl quantization 
rule~(\ref{eq:WeylMap2}) the product of two 
field operators defined at two causally connected 
spacetime points
$x_1\in\Sigma_1$ and $x_2\in\Sigma_2$  
\beq
\label{eq:prodPhis}
& & \hat{\Phi}^{I}(x_{1})  \hat{\Phi}^{J}(x_{2})=
W\left[\Phi^{I}(x_{1})\right]W\left[\Phi^{J}(x_{2}) \right] \nonumber\\
& = &    W\left[ -\int_{\Sigma_{2}}\left( \tilde{G}^{IK}(x_{1},x'_{2})\frac{\partial\Phi^{K}(x'_{2})}{\partial x^{0}_{2}}-\Phi^{K}(x'_{2})\frac{\partial\tilde{G}^{IK}(x_{1},x'_{2})}{\partial{x_{2}^{0}}}\right)d\Sigma_{2}\right] 
W\left[\Phi^{J}(x_{2}) \right] \,,\nn\\
\eeq
where $x'_2$ also belongs to the hypersurface $\Sigma_2$.  In this way, formula~(\ref{eq:prodPhis}) relates field operators at two different hypersurfaces, $\Sigma_1$ and $\Sigma_2$, by representing the operator at 
hypersurface $\Sigma_1$ by a corresponding operator defined at the hypersurface $\Sigma_2$ through the causal 
Green's function.  
In order to use the Moyal product~(\ref{eq:MoyalTr}), in 
our case we consider the functions  
\beq
F_1 & := &  -\int_{\Sigma_{2}}\left( \tilde{G}^{IK}(x_{1},x'_{2})\frac{\partial\Phi^{K}(x'_{2})}{\partial x^{0}_{2}}-\Phi^{K}(x'_{2})\frac{\partial\tilde{G}^{IK}(x_{1},x'_{2})}{\partial{x_{2}^{0}}}\right)d\Sigma_{2}  \,,\nn\\
F_2 & := & \Phi^{J}(x_{2})  \,.
\eeq 
By means of the properties of the Wigner 
function~(\ref{eq:MoyalTr})
the 
star-product
reads
\begin{eqnarray}
& & \Phi^{I}(x_{1})\star\Phi^{J}(x_{2})
= 
W^{-1}\left[ W(\Phi^{I}(x_{1}))W(\Phi^{J}(x_{2}))\right]
= 
W^{-1}\left[ W(F_1)W(F_2)\right] \nn\\
& = &
\Tr \left\lbrace  \int\mathcal{D}\Phi'\mathcal{D}\left( \frac{\Pi'}{2\pi\hbar}\right) \mathcal{D}\Phi''\mathcal{D}\left(\frac{\Pi''}{2\pi\hbar}\right) \hat{\Omega}(\Phi,\Pi)\hat{\Omega}(\Phi',\Pi')\hat{\Omega}(\Phi'',\Pi'') \right.  \nonumber \\
&& \times \left.\left[ -\int_{\Sigma_{2}}\left( \tilde{G}^{IK}(x_{1},x'_{2})
\frac{\partial{\Phi}'^{K}(x'_2)}{\partial x_2^0}-\Phi'^{K}(x'_2)\frac{\partial\tilde{G}^{IK}(x_{1},x'_{2})}{\partial{x_{2}^{0}}}\right)d\Sigma_{2}\right]
\Phi''^{J}(x_2) \right\rbrace \,.\nn\\
\label{eq:MoyalKirc}
\end{eqnarray}
Here the $\Omega$'s stand for the Stratonovich-Weyl quantizer~(\ref{eq:StratoW}), one resulting 
directly from the definition of the 
star-product~(\ref{eq:MoyalTr}), and the other
two coming from the Weyl quantization 
rule~(\ref{eq:WeylMap2}).   
As discussed in~\cite{Compean}, considering the trace properties of  the 
Stratonovich-Weyl quantizer~(\ref{eq:SW1})-(\ref{eq:SW3})
we may write the star-product in its integral representation
\beq
\label{eq:MoyalProdSW}
& & 
\Phi^{I}(x_{1})\star\Phi^{J}(x_{2}) \nn\\
&=& \int\mathcal{D}\Phi'\mathcal{D}\left( \frac{\Pi'}{2\pi\hbar}\right) \mathcal{D}\Phi''\mathcal{D}\left(\frac{\Pi''}{2\pi\hbar}\right)    \nonumber \\
&& \times \left[ -\int_{\Sigma_{2}}\left( \tilde{G}^{IK}(x_{1},x'_{2})
\frac{\partial{\Phi}'^{K}(x'_2)}{\partial x_2^0}-\Phi'^{K}(x'_2)\frac{\partial\tilde{G}^{IK}(x_{1},x'_{2})}{\partial{x_{2}^{0}}}\right)d\Sigma_{2}\right]
\Phi''^{J}(x_2)  \nn\\
& & 
\times\exp \left\lbrace \frac{2i}{\hbar}\int dx \left( (\Phi-\Phi')\cdot(\Pi-\Pi'')-(\Phi-\Phi'')\cdot(\Pi-\Pi')\right) \right\rbrace
\eeq
We must emphasize that we have used the representation~(\ref{eq:Kirc}) to map both field operators to operators defined at different points in the same hypersurface for a fixed time parameter,
that is, $x_2,x'_2\in \Sigma_2$. 
Bearing in mind this, from now on, we will 
avoid the primes for points in the same hypersurface 
($x_2,x'_2\in\Sigma_2$ before), and 
thus, making a small abuse of language, we will 
refer to points in the same hypersurface with the same 
symbol.
Finally, Taylor expanding~(\ref{eq:MoyalProdSW}) with respect
to the field variables we see 
that, after tedious but straightforward calculations, expression~(\ref{eq:MoyalKirc}) is 
reduced to 
\begin{equation}
\Phi^{I}(x_{1})\star\Phi^{J}(x_{2})=\Phi^{I}(x_{1})\Phi^{J}(x_{2})+\frac{i\hbar}{2}\tilde{G}^{IJ}(x_{1},x_{2}) \,.
\end{equation}
This result encompass the general behaviour of the Moyal product, and may be used to define the star-commutator 
\begin{equation}
\left[\Phi^{I}(x_{1}),\Phi^{J}(x_{2}) \right]:=\Phi^{I}(x_{1})\star\Phi^{J}(x_{2})- \Phi^{J}(x_{2})\star\Phi^{I}(x_{1})
\,,
\end{equation}
which in our case simply reduces to 
\beq
\label{eq:MoyalComm}
\left[\Phi^{I}(x_{1}),\Phi^{J}(x_{2}) \right]= i\hbar\tilde{G}^{IJ}(x_{1},x_{2})  \,.
\eeq
Of course, the 
star-commutator~(\ref{eq:MoyalComm})
must follow the deformation quantization 
axioms (i) to (iii) of Section~\ref{sec:review}.
Properties (i) and (ii) are directly satisfied.  
However, property (iii), the so-called correspondence principle, indicates that 
this star-commutator is interrelated to 
a classical Poisson structure at 
two different spacetime points given by
\begin{equation}
\label{eq:CovProd}
\left( \Phi^{I}(x_{1}),\Phi^{J}(x_{2})\right)=\tilde{G}^{IJ}(x_{1},x_{2}) \,. 
\end{equation}
Here we used round brackets instead of curly 
brackets in order to make a distinction from the 
standard Poisson bracket.  It is important to 
mention that our classical functional bracket~(\ref{eq:CovProd}) results, by 
construction, consistent with the product of 
two field operators at different spacetime points.  
Further, in analogy with the calculation above, it is straightforward to generalize our 
classical functional bracket to arbitrary functionals $F_1(\Phi(x_1)), \ F_2(\Phi(x_2))$ of the fields variables at two causally connected points
$x_1,x_2\in\mathcal{M}$ by the relation
\begin{equation}
F_{1}[\Phi(x_{1})]\star F_{2}[\Phi(x_{2})]=F_{1}[\Phi(x_{1})]\exp\left\lbrace\frac{i\hbar}{2}\stackrel{\leftrightarrow}{\mathcal{K}} \right\rbrace  F_{2}[\Phi(x_{2})]  \,, 
\end{equation}
where the bidifferential operator 
$\stackrel{\leftrightarrow}{\mathcal{K}}$
is explicitly given by 
\begin{equation}
\stackrel{\leftrightarrow}{\mathcal{K}}:=\exp \left\lbrace \frac{i\hbar}{2}\int dxdx'\left( \frac{\overleftarrow{\delta}\ \ \ \ \ }{\delta\Phi^{M}(x)}\tilde{G}^{MN}(x,x')\frac{\overrightarrow{\delta}\ \ \ \ \ }{\delta\Phi^{N}(x')}\right)  \right\rbrace \,.
\end{equation}
Once again, by the correspondence principle (iii), this star-product leads to a well-defined Poisson structure given by
\begin{equation}
\hspace{-5ex}
\label{eq:CovBracket}
(F_{1}[\Phi(x_{1})],F_{1}[\Phi(x_{2})]):=\int_{\mathcal{M}} dxdx'\frac{\delta F_{1}[\Phi(x_{1})]}{\delta\Phi^{M}(x)}\tilde{G}^{MN}(x,x')\frac{\delta F_{2}[\Phi(x_{2})]}{\delta\Phi^{N}(x')} \,,
\end{equation}
which, as it is expected, turns out to be 
skew-symmetric, bilinear and obey both, 
the Jacobi identity~(\ref{eq:Jacobi}) and the 
Leibniz condition~(\ref{eq:Leibniz}).  
Bearing in mind ~(\ref{eq:Kirc}), we must emphasize 
the different manners in which the bracket~(\ref{eq:CovBracket}) must be understood:
for linear systems the Green's function gives us an integral 
representation of the solution while for nonlinear systems  the associated Green's function 
approximately characterizes solutions only in an
appropriate domain and depends on specific boundary 
conditions inherent to a given 
system\footnote{The method to solve nonlinear differential 
equations by means of Green's functions may be shown to be equivalent to a small expansion in the 
spacetime parameters, even though for some cases
we need to consider slow convergence
as demonstrated in~\cite{Frasca1,Frasca2}}.   
The fact that the 
functional bracket~(\ref{eq:CovBracket}) is
indeed a causal Poisson structure following these
properties is shown in~\ref{sec:MathProps}. 
It should also be noted that the causal
Poisson bracket~(\ref{eq:CovBracket}) may be extended 
naturally to include gradients of the fields
\beq
\left(\partial_{\mu}\Phi^{I}(x),\Phi^{J}(x') \right) & = & \partial_{\mu}\tilde{G}^{IJ}(x,x')\,, \nn\\
\left(\Phi^{I}(x),\partial_{\mu'}\Phi^{J}(x') \right) & = & \partial_{\mu'}\tilde{G}^{IJ}(x,x')\,,\nn\\
\left(\partial_{\mu}\Phi^{I}(x),\partial_{\nu'}\Phi^{J}(x') \right)& = &\partial^{2}_{\mu\nu'}\tilde{G}^{IJ}(x,x')\,.
\eeq
which, in turn, by considering the standard 
momentum definition may be extended to the phase
space associated to the fields $\Phi$, that we will 
denote as $\Gamma(\Phi,\Pi)$.  Besides, when 
$\Phi^{I}(x)$ and $\Phi^{J}(x')$ are defined in the same hypersurface $\Sigma$, that is, whenever $x^{0}$ tends to $x'^{0}$, we recover the 
standard field theoretic Poisson structure.   We 
may deduce this directly from the discontinuity properties of the causal Green's function $\tilde{G}^{IJ}$, since this 
function and its derivatives follow the limits
\beq
\lim_{x^{0}\rightarrow x'^{0}}\tilde{G}^{IJ}(x,x') & = & 
0  \,,\nn\\
\lim_{x^{0}\rightarrow x'^{0}}\partial_{x^{0}}\tilde{G}^{IJ}(x,x') & = & 
\delta^{3}(\vec{x},\vec{x}')  \,,\nn\\
\lim_{x^{0}\rightarrow x'^{0}}\partial^{2}_{x^{0}x'^{0}}\tilde{G}^{IJ}(x,x') 
& = & 0  \,.
\eeq
The first limit holds from to the definition 
of the causal Green's function in terms of the 
advanced and retarded Green's functions.
The second limit simply states the discontinuity 
of the causal Green's function.  Finally, 
the third limit holds since the second derivative of both $G^{+IJ}$ and $G^{-IJ}$ are proportional to a Dirac delta distribution.

For the case of theories involving interacting fields we are confined  to a perturbative framework, then we are interested in the star product of $n$ fields, where due to the properties of the causal Poisson bracket, and the combinatorics of all contractions, this product becomes into a generalized version of the Wick's 
theorem
\beq
& & 
\Phi^{I_{1}}(x_{1})  \star \Phi^{I_{2}}(x_{2})\star  \cdots\star\Phi^{I_{n}}(x_{n})
\nn\\
& = &
\Phi^{I_{1}}(x_{1})\Phi^{I_{2}}(x_{2})\cdots\Phi^{I_{n}}(x_{n}) 
\nn \\
& & 
+\left(\frac{i\hbar}{2}\right) 
\mathop{\sum_{\mathtt{single}}}_{\mathtt{pairs}}\left[\tilde{G}^{I_{i}I_{j}}(x_{i},x_{j})\Phi^{I_{1}}(x_{1})\cdots\widehat{\Phi^{I_{i}}(x_{i})}\cdots\widehat{\Phi^{I_{j}}(x_{j})}\cdots \Phi^{I_{n}}(x_{n})  \right]
\nn\\
& & 
+\left( \frac{i\hbar}{2}\right)^{2} 
\mathop{\sum_{\mathtt{double}}}_{\mathtt{pairs}}\left[\tilde{G}^{I_{i}I_{j}}(x_{i},x_{j}){G}^{I_{k}I_{l}}(x_{k},x_{l})\Phi^{I_{1}}(x_{1})\cdots\widehat{\Phi^{I_{i}}(x_{i})}\cdots\widehat{\Phi^{I_{j}}(x_{j})}\cdots \right.\nn\\
& & \left.\cdots\widehat{\Phi^{I_{k}}(x_{k})}\cdots\widehat{\Phi^{I_{l}}(x_{l})}\cdots \Phi^{I_{n}}(x_{n})\right] 
+\cdots 
\eeq
where $\widehat{\Phi^{I_{i}}(x_{i})}$ denotes that the field $\Phi^{I_{i}}(x_{i})$ has been removed from the summation. The first sum runs over single contractions of pairs, while the second sum runs over double contractions, and so on. If $n$ is even, the product ends with terms only consisting of products of casual Green's functions.
By making use of the equivalence of star products stated by means 
of isomorphisms between star algebras in~(\ref{eq:StarEquiv}), it %
is possible to write the time ordered product of quantum field 
operators through the normal ordering 
map $\Theta_{N}$~\cite{Dito1}.  This normal ordering map 
sends any functional $F$ defined on the phase space 
to the associated normal ordering operator $
\Theta_{N}[F]$. Then 
\begin{eqnarray}
T\left\lbrace\hat{\Phi}(x_{1})\cdots\hat{\Phi}(x_{n})\right
\rbrace  
& = &  
\Theta_{N}\left\lbrace \exp \left[  \frac{i\hbar}{2}
\int dxdx'\left( \frac{\overleftarrow{\delta}}{\delta\Phi^{M}(x)}
G^{MN}_{F}(x,x')\frac{\overrightarrow{\delta}}{\delta\Phi^{N}
(x')}\right) \right] \right.
\nn\\
& & 
\left.\times\prod_{p=1}^{n}\Phi^{I_{p}}(x_{p})\right\rbrace
\,. 
\end{eqnarray}    
Here, $G_{F}^{MN}(x,x')$ stands for the Feynman 
propagator.
We can observe that the time ordered product $T$
do not correspond to the Weyl transform of a causal star-product %
since, by definition, the time ordered product is fully symmetric 
in its arguments while the causal 
star-product have skew-symmetry properties 
inherited from the construction of the causal Green's function. 
Further, we see that the 
causal and Feynman Green's functions may be constructed in terms 
of different combination of primitive Green's functions as
\begin{eqnarray}
\tilde{G}&=& G^{+}-G^{-}=G^{(+)}+G^{(-)} \,, 
\nn\\
G_{F}&=& G^{-}+G^{(-)}=G^{+}-G^{(+)}  \,,
\end{eqnarray}
where $G^{+}$ and $G^{-}$ are the advanced and retarded 
Green's functions, whereas $G^{(+)}$ and $G^{(-)}$ correspond to 
the positive and negative frequency propagators~\cite{Dewitt1}. 
By the preceding arguments,  Wick's theorem could also be written as a generating function
\begin{eqnarray}
\hspace{-2ex}
\label{Wick}
T \left\lbrace \exp \left[ \frac{i}{\hbar}\int d^{4}x\, J(x)\hat{\Phi}(x)\right] \right\rbrace & = &  \Theta_{N}\left\lbrace \exp\left( \frac{i}{\hbar}\right)\int d^{4}x J(x)\hat{\Phi}(x)\right\rbrace 
\nn \\ 
& &  \exp\left[ -\frac{1}{2\hbar^{2}}\int d^{4}xd^{4}x'\,J(x)G_{F}(x,x')J(x') \right]\,, \nn\\
\end{eqnarray}     
where $J(x)$ denotes an external source~\cite{Hirshfeld1}. Expanding  equation~(\ref{Wick}) in powers of $J$, we note that this term corresponds to the perturbation expansion of the scattering operator in quantum field theory, which has been derived entirely under the deformation quantization framework.

%
%

\section{Examples}
\label{sec:Examples}

In this section we put our previously obtained results  
at work by 
exploring some specific physically motivated examples .

\subsection{Real scalar field}

We will work on Minkowski spacetime $\mathcal{M}$. 
The action for a real scalar field $\phi:\mathcal{M}\rightarrow \mathbb{R}$ reads
\beq
S_{\mathrm{KG}}[\phi] = -\int_{\mathcal{M}} d^{4}x \ \frac{1}{2}\left[ \left(\partial^{\mu}\phi\right)\left(\partial_{\mu}\phi\right)-m^{2}\phi^{2}\right]   \,,
\eeq
where $\mu=0,1,2,3$ denote spacetime indices, and $m$ is a 
constant mass term.  Motion of the field is given by the 
well-known Klein-Gordon  equation
\beq
\left(\partial_{\mu}\partial^{\mu}+m^{2}\right)\phi=0
\label{eq:kleingordon}
\eeq
for which we may associate the usual advanced and retarded 
Green's functions
\beq
G^{+}(x,y) & = & \frac{-i}{(2\pi)^{3}}\int \frac{d^{3}\vsp{k}}{2\omega(\vsp{k})}e^{i\left(\omega(\vsp{k})(x^{0}-y^{0})-\vsp{k}(\vsp{x}-\vsp{y})\right)} \,, 
\hspace{4ex}\\
G^{-}(x,y) & = & \frac{-i}{(2\pi)^{3}}\int \frac{d^{3}\vsp{k}}{2\omega(\vsp{k})}e^{i\left(-\omega(\vsp{k})(x^{0}-y^{0})-\vsp{k}(\vsp{x}-\vsp{y})\right)} ,
\hspace{4ex}
\eeq
respectively.  Here we have written $\omega(\vsp{k})=\pm\sqrt{\vsp{k}^{2}+m^{2}}$.  Given two spacetime points 
$x,y\in\mathcal{M}$, by relation~(\ref{eq:CausalGreen})
we construct  the causal Green's function $\tilde{G}(x,y)$ as
\beq
\tilde{G}(x,y) = \frac{1}{(2\pi)^{3}}\int \frac{d^{3}\vsp{k}}{\omega(\vsp{k})}\sin{\left[\omega(\vsp{k})(x^{0}-y^{0})\right]}e^{-i\vsp{k}(\vsp{x}-\vsp{y})}  \,. \nn\\
 \label{eq:causalgrensca}
\eeq
It may be easily shown that this causal Green's function follows 
the equal time limits
\beq
\lim_{x^{0}\rightarrow y^{0}}\tilde{G}(x,y) & = & 0 \, , \nn \\
\lim_{x^{0}\rightarrow y^{0}}\frac{\partial \tilde{G}(x,y)}{\partial x^{0}} & = & \delta(\vsp{x}-\vsp{y}) =
-\lim_{x^{0}\rightarrow y^{0}}\frac{\partial \tilde{G}(x,y)}{\partial y^{0}}  \, , \nn \\
\lim_{x^{0}\rightarrow y^{0}}\frac{\partial^{2} \tilde{G}(x,y)}{\partial x^{0}\partial y^{0}} & = & 0 \, ,
\label{eq:limitesgreen}
\eeq
which are relevant in our formulation in order to recover
in this limit the 
standard Poisson bracket for the field $\phi(x)$ and its
conjugate momentum $\pi(x):=(\partial \mathcal{L}/\partial 
\dot{\phi})(x)=\dot{\phi}(x)$, where the dot means derivative with respect
to the $x_0$ parameter.  Thus, the causal Green's function 
$\tilde{G}(x,y)$ 
is used  to establish the integral representation
\beq
\phi(x) & = &  -\int_{\Sigma}\left(\tilde{G}(x,y)\pi(y)-\phi(y)\frac{\partial \tilde{G}(x,y)}{\partial y^{0}}\right)d^{3}y
\hspace{4ex} \nn \\
\pi(x) & = & -\int_{\Sigma}\left(\frac{\partial \tilde{G}(x,y)}{\partial x^{0}}\pi(y)-\phi(y)\frac{\partial^{2} \tilde{G}(x,y)}{\partial x^{0}\partial y^{0}}\right)d^{3}y ,
\hspace{4ex}
\label{eq:flujogreenesc}
\eeq
where integrals are taken over a given hypersurface $\Sigma$.
From this representation, and by considering the causal Poisson 
brackets introduced in~(\ref{eq:CovBracket}), we find the elementary 
causal brackets
\beq
(\phi(x),\phi(y)) & = & \tilde{G}(x,y) \, , \nn \\
(\phi(x),\pi(y))  & = & \frac{\partial \tilde{G}(x,y)}{\partial x^{ 0}} \, , \nn \\
(\pi(x),\phi(y))  & = & \frac{\partial \tilde{G}(x,y)}{\partial y^{ 0}} \, , \nn \\
(\pi(x),\pi(y))   & = & \frac{\partial^{2} \tilde{G}(x,y)}{\partial x^{0}\partial y^{0}} \, .
\label{eq:poiesca}
\eeq
As stated before, by considering the limits
(\ref{eq:limitesgreen}) we see that these 
causal brackets are simplified to the standard equal-time 
classical Poisson brackets at two different spatial points on a given hypersurface 
$\Sigma$.

Next, the real scalar field, $\phi(x)$, and 
its conjugate momentum, $\pi(x)$, may be written
in terms of the annihilation and creation coefficients,  $a(\vsp{k},x^{0})$ and 
$a^{*}(\vsp{k},x^{0})$, respectively, as 
\beq
\label{eq:realphi}
\phi(x)=\frac{1}{(2\pi)^{3}}\int d^{3}\vsp{k}\left(\frac{\hbar}{2\omega(\vsp{k})}\right)^{1/2}\left(a(\vsp{k},x^{0})e^{i\vsp{k}\vsp{x}}+a^{*}(\vsp{k},x^{0})e^{-i\vsp{k}\vsp{x}}\right) \,,
\\
\label{eq:realpi}
\pi(x)=\frac{i}{(2\pi)^{3}}\int d^{3}\vsp{k}\left(\frac{\hbar\omega(\vsp{k})}{2}\right)^{1/2}\left(-a(\vsp{k},x^{0})e^{i\vsp{k}\vsp{x}}+a^{*}(\vsp{k},x^{0})e^{-i\vsp{k}\vsp{x}}\right) \,,
\eeq
where $a(\vsp{k},x^{0}):=a(\vsp{k})e^{-i\omega(\vsp{k})x^{0}}$.  As usual,
relations~(\ref{eq:realphi})
and~(\ref{eq:realpi}) may be inverted in order to 
find the coefficients
\beq
a(\vsp{k},x^{0}) & = & \frac{1}{(2\hbar\omega(\vsp{k}))^{1/2}}\int d^{3}\vsp{x}e^{-i\vsp{k}\vsp{x}}\left(\omega(\vsp{k})\phi(x)+i\pi(x)\right) \,,\nn \\
a^{*}(\vsp{k},x^{0}) & = & \frac{1}{(2\hbar\omega(\vsp{k}))^{1/2}}\int d^{3}\vsp{x}e^{i\vsp{k}\vsp{x}}\left(\omega(\vsp{k})\phi(x)-i\pi(x)\right)  \,.
\eeq
By repeatedly applying the causal bracket
relations~(\ref{eq:poiesca}) we may obtain 
the classical commutation rules
\beq
\label{eq:poia1}
(a(\mathbf{k},x^{0}),a^{*}(\mathbf{k}',y^{0})) & = & \frac{i\left(\omega(\vsp{k})+\omega(\vsp{k}')\right)}{\left(4\omega(\vsp{k})\omega(\vsp{k'})\right)^{1/2}\hbar}e^{i\omega(\vsp{k})(x^{0}-y^{0})}\delta(\vsp{k}-\vsp{k}') \, ,  \\
\label{eq:poia2}
(a(\mathbf{k},x^{0}),a(\mathbf{k}',y^{0})) & = & -\frac{i\left(\omega(\vsp{k})-\omega(\vsp{k}')\right)}{\left(4\omega(\vsp{k})\omega(\vsp{k'})\right)^{1/2}\hbar}e^{i\omega(\vsp{k})(x^{0}-y^{0})}\delta(\vsp{k}-\vsp{k}')  \, ,  \\
(a^{*}(\mathbf{k},x^{0}),a^{*}(\mathbf{k}',y^{0})) & = & \frac{i\left(\omega(\vsp{k})-\omega(\vsp{k}')\right)}{\left(4\omega(\vsp{k})\omega(\vsp{k'})\right)^{1/2}\hbar}e^{-i\omega(\vsp{k})(x^{0}-y^{0})}\delta(\vsp{k}-\vsp{k}') \,,  
 \label{eq:poia3}
\eeq
where we have substituted the causal Green's function~(\ref{eq:causalgrensca}) and its 
derivatives, and we have explicitly performed
the involved integrals.  From these 
classical commutators
we note that, due to the $\omega(\vsp{k})-\omega(\vsp{k}')$ factor and to the Dirac 
delta $\delta(\vsp{k}-\vsp{k}')$, the 
last two brackets are vanishing in 
a distributional sense, indicating the 
non-interacting nature of the annihilation and creation coefficients at different times.  
This may be interpreted as a manifestation of 
energy conservation at two different spatial 
hypersurfaces.  Also, the first classical commutation rule~(\ref{eq:poia1}) generalizes 
the standard Poisson bracket allowing annihilation and creation coefficients at different spatial hypersurfaces.  Clearly,
these classical commutation rules reduce 
to the standard Poisson bracket at the 
equal-time limit.

This example may be also relevant for the 
analysis of the free electromagnetic field
given by the action 
\beq
S_{\mathrm{EM}}[A]  :=  -\frac{1}{4}\int d^{4}x \ F_{\mu\nu}F^{\mu\nu}
\eeq
where the electromagnetic field $F_{\mu\nu}$ may be written in a common way in terms of the 
potential vector field $A(x)$ by the relation $F_{\mu\nu}:=\partial_{\mu}A_{\nu}-\partial_{\nu}A_{\mu}$ ($\mu,\nu=0,1,2,3$).  Indeed, by fixing 
the radiation gauge $A^0=0$, for example, the 
field equations for the spatial components 
reduce to $\partial_{\mu}\partial^{\mu}A^{i} = 0$ 
which may be thought of as a non-massive Klein-Gordon equation for each of the spatial components $A^i$.  In this sense, 
the electromagnetic case may be interpreted
as three independent real scalar fields, 
as formulated in~\cite{Compean}, and thus we 
may, in principle, extrapolate the results 
obtained here to analyze the electromagnetic field.

\subsection{Bosonic string theory}

As it is well known, the relativistic boson string may be 
described by the Nambu-Goto action.  Variation of this action 
leads to non-linear equations of motion for the string 
due to the complexity of the momenta involved.  In order to avoid 
this issues, we will then start with the classically equivalent Polyakov 
action~\cite{polchi,BBS,CompeanBosonic}
\beq
S_\mathrm{P}[X]= -\frac{1}{4\pi\alpha'}\int_\Sigma d\tau d\sigma(-\gamma)^{1/2}\gamma^{ab}\partial_{a}X^{\mu}\partial_{b}X^{\nu}\eta_{\mu\nu}  \,.
\eeq
Here the world-sheet $\Sigma$ swept out by the string 
is parametrized by $(\sigma,\tau)$, and $X^\mu$ is the 
embedding of the world-sheet $\Sigma$ into the 
spacetime manifold  $\mathcal{M}$ endowed with a 
metric $\gamma^{ab}(\sigma,\tau)$ ($\gamma=
\det(\gamma_{ab})$).
Finally, 
$\alpha'$ is a parameter associated to  the string scale 
squared, and may be commonly thought of as proportional to the 
inverse of the string tension.   Equations of motion for the 
Polyakov action may be substantially reduced if one considers 
the choice $\gamma_{ab}  =  \eta_{ab}e^{\phi}$, where 
$\eta$ is a two-dimensional Minkowski metric and $e^\phi$ 
is the conformal factor for the spacetime function $\phi$. 
In this way, the equations of motion are simply given by 
\beq
\left(\frac{\partial^{2}}{\partial \sigma^{2}}-\frac{\partial^{2}}{\partial \tau^{2}}\right)X^{\mu}=0 \,,
\label{eq:ecmovpo}
\eeq
that is, the two-dimensional wave equation.

In order to construct the causal Green's function, $\tilde{G}(\sigma,
\sigma';\tau,\tau')$, we need first to impose appropriate 
boundary conditions: For an open string the total derivative
term $\partial_\nu X^\mu$ 
fix the boundary contributions, while  
periodicity conditions on the parameter $\sigma$ must also 
be considered for a closed string. In this sense, we will consider boundary conditions $X'^{\mu}(\tau,0) =  0 = X'^{\mu}(\tau,\pi)$ for the open string, $X'^{\mu}(\tau,-\infty) =  0 = X'^{\mu}(\tau,\infty)$
for the infinitely open string, 
and $X^{\mu}(\tau,0) =  X^{\mu}(\tau,\pi),\ X'^{\mu}(\tau,0) =  X'^{\mu}(\tau,\pi)$ together with $\gamma^{ab}(\tau,0) =  \gamma^{ab}(\tau,\pi)$, for the closed string, respectively. 
Here $X'^\mu(\sigma,
\tau)$ denotes derivative with respect to the parameter $\sigma$.
Thus, depending on these boundary conditions, we may find 
in a complete standard manner the 
causal Green's functions
\beq
\label{eq:causalabierta}
\tilde{G}_{\mathrm{open}}(\sigma,\sigma';\tau,\tau') & = &\sum_{n}\frac{1}{2n}\sin{2n(\tau-\tau')}\cos{2n(\sigma-\sigma')} \,,\\
\label{eq:causalinfty}
\tilde{G}_{\infty}(\sigma,\sigma';\tau,\tau') & = & \theta{\left[(\tau-\tau')-(\sigma-\sigma')\right]} \,,\\
\label{eq:causalcerrada}
\tilde{G}_{\mathrm{closed}}(\sigma,\sigma';\tau,\tau') & = &\sum_{n}\frac{1}{2n}\sin{2n(\tau-\tau')}\cos{2n\sigma}\cos{2n\sigma'} \,,
\eeq
for the open, the infinitely open, and closed strings, 
respectively~\cite{Das}.  In~(\ref{eq:causalinfty}), the 
$\theta$ stands for the Heaviside step-function.
It is easy to see that for the three 
cases~(\ref{eq:causalabierta})-(\ref{eq:causalcerrada}), the corresponding 
causal Green's functions follow the limits
\beq
\lim_{\tau\rightarrow\tau'}\tilde{G}(\sigma,\sigma';\tau,\tau') & = & 0 \, , 
\nn \\
\lim_{\tau\rightarrow\tau'}\frac{\partial\tilde{G}(\sigma,\sigma';\tau,\tau')}{\partial \tau} & = & \delta(\sigma-\sigma')\,, 
\nn\\
\lim_{\tau\rightarrow\tau'}\frac{\partial\tilde{G}(\sigma,\sigma';\tau,\tau')}{\partial \tau'} & = & -\delta(\sigma-\sigma')\, , 
\nn \\
\lim_{\tau\rightarrow\tau'}\frac{\partial^{2}\tilde{G}(\sigma,\sigma';\tau,\tau')}{\partial \tau\partial \tau'} & = & 0 \,,
\label{eq:limGnambu}
\eeq
which are very important in order to 
guarantee the standard Poisson bracket limit, as stated in 
Section~\ref{sec:CovPoissonStructure}.

By considering the associated momenta to the fields $X^\mu$, 
$\Pi^{\mu}:= \partial \mathcal{L}/\partial (\dot{X}^{\mu}) = (2\pi \alpha')^{-1}\dot{X}^{\mu}$, where the dot means derivative with respect to the 
parameter $\tau$, and by using the integral representation~(\ref{eq:Kirc}) we obtain
\beq
X^{\mu}(\sigma,\tau) & = & -\int d\sigma'\left(2\pi\alpha'\tilde{G}(\sigma,\sigma';\tau,\tau')\Pi^{\mu}(\sigma',\tau')\right. \nn\\
 & & \left.-\frac{\partial \tilde{G}(\sigma,\sigma';\tau,\tau')}{\partial \tau'}X^{\mu}(\sigma',\tau')\right) \nn \\
\Pi^{\mu}(\sigma,\tau) & = & -\frac{1}{2\pi\alpha'}\int d\sigma'\left(2\pi\alpha'\frac{\partial \tilde{G}(\sigma,\sigma';\tau,\tau')}{\partial \tau}\Pi^{\mu}(\sigma',\tau') \right. \nn \\
 & & \left.-\frac{\partial^{2} \tilde{G}(\sigma,\sigma';\tau,\tau')}{\partial \tau \partial \tau'}X^{\mu}(\sigma',\tau')\right) \,,
\eeq 
for any of the causal Green's functions~(\ref{eq:causalabierta})-(\ref{eq:causalcerrada}).  Thus, for these variables we construct the fundamental causal brackets 
\beq
(X^{\mu}(\sigma,\tau),X^{\nu}(\sigma',\tau')) & = & 2\pi\alpha'\eta^{\mu\nu}\tilde{G}(\sigma,\sigma';\tau,\tau') \,, 
\nn \\
(X^{\mu}(\sigma,\tau),\Pi^{\nu}(\sigma',\tau')) & = & \eta^{\mu\nu}\frac{\partial \tilde{G}(\sigma,\sigma';\tau,\tau')}{\partial \tau} 
\,, 
\nn \\
(\Pi^{\mu}(\sigma,\tau),X^{\nu}(\sigma',\tau')) & = & \eta^{\mu\nu}\frac{\partial \tilde{G}(\sigma,\sigma';\tau,\tau')}{\partial \tau'} \,,
\nn \\
(\Pi^{\mu}(\sigma,\tau),\Pi^{\nu}(\sigma',\tau')) & = & \frac{\eta^{\mu\nu}}{2\pi\alpha'}\frac{\partial^{2} \tilde{G}(\sigma,\sigma';\tau,\tau')}{\partial \tau \partial \tau'}  \,.
\label{eq:poinambu} 
\eeq
As discussed before, these brackets reduce in the equal-time 
limit to the standard Poisson brackets.  Furthermore, 
the general solution to the wave equation~(\ref{eq:ecmovpo}) is given by 
$X^{\mu}(\sigma,\tau) = X_{L}^{\mu}(\tau+\sigma)+X_{R}^{\mu}(\tau-\sigma)$, for which we may write explicitly
\beq
X_{L}^{\mu}(\tau+\sigma) & = & \frac{1}{2}x^{\mu}+\alpha' \pi^{\mu}(\tau+\sigma)+i\left(\frac{\alpha'}{2}\right)^{1/2}\sum_{n\neq 0}\frac{1}{n}\tilde{\alpha}_{n}^{\mu}e^{-2in(\tau+\sigma)} \, , \nn \\
X_{R}^{\mu}(\tau-\sigma) & = & \frac{1}{2}x^{\mu}+\alpha' \pi^{\mu}(\tau-\sigma)+i\left(\frac{\alpha'}{2}\right)^{1/2}\sum_{n\neq 0}\frac{1}{n}\alpha_{n}^{\mu}e^{-2in(\tau-\sigma)} \,,
\label{eq:solucuerdaabierta}
\eeq
that is, we may Fourier expand in terms of left or right moving
oscillation modes, respectively.\footnote{This solution stands for the open
string.  For the closed string we may also consider the 
relations $\alpha^{\mu}_{n}=\tilde{\alpha}^{\mu}_{n}$ in order to 
preserve the appropriate boundary conditions.}  Here the coefficients $x^{\mu}$  and 
$\pi^{\mu}$ correspond to the $n=0$ expansion terms, 
and are defined in terms of $X^{\mu}$ and $\Pi^{\mu}$ by the relations
\beq
x^{\mu} & := & \frac{1}{2\pi}\int X^{\mu}(\sigma,0)d\sigma \,, 
\nn \\
\pi^\mu & := & \int \Pi^{\mu}(\sigma,0)d\sigma \,. 
\eeq
Also, in a standard manner we find the coefficients
$\alpha_{n}^{\mu}(\tau):= e^{-2in\tau}\alpha_{n}^{\mu}$ and 
$\tilde{\alpha}_{n}^{\mu}(\tau) := e^{2in\tau}\tilde{\alpha}_{n}^{\mu}$ in terms of $X^{\mu}$ and $\Pi^{\mu}$ by
\beq
\alpha_{n}^{\mu}(\tau) & = & \left(\frac{2}{\alpha'}\right)^{1/2}\int \left(-\frac{in}{2\pi}X^{\mu}(\sigma,\tau)+\frac{\alpha'}{2}\Pi^{\mu}(\sigma,\tau)\right)e^{-2in\sigma}d\sigma \nn \\
\tilde{\alpha}_{n}^{\mu}(\tau) & = & \left(\frac{2}{\alpha'}\right)^{1/2}\int \left(-\frac{in}{2\pi}X^{\mu}(\sigma,\tau)+\frac{\alpha'}{2}\Pi^{\mu}(\sigma,\tau)\right)e^{2in\sigma}d\sigma \, .
\label{eq:anicreanambu}
\eeq
As it is expected, coefficients $\alpha^\mu_n$ and 
$\tilde{\alpha}^\mu_n$ are related to the creation and 
annihilation coefficients.  Using the causal Green's function
for an open string~(\ref{eq:causalabierta}), and by
repeated application of brackets~(\ref{eq:poinambu}), we are 
able to evaluate the causal brackets for these coefficients
\beq
(\alpha_{n}^{\mu}(\tau),\alpha_{m}^{\nu}(\tau')) & = & -im\delta_{n+m}e^{-2in(\tau-\tau')}\eta^{\mu\nu} \, , \nn \\
(\tilde{\alpha}_{n}^{\mu}(\tau),\tilde{\alpha}_{m}^{\nu}(\tau')) & = & -im\delta_{n+m}e^{-2in(\tau-\tau')}\eta^{\mu\nu} \, , \nn \\
(\alpha_{n}^{\mu}(\tau),\tilde{\alpha}_{m}^{\nu}(\tau')) & = & m\sin{2n(\tau-\tau')}\delta_{nm}\eta^{\mu\nu} \, .
\label{eq:anicreaNambu}
\eeq
Note that these relations reduce to the standard relations in the equal-time limit. In addition, we may define the familiar classical observables
\beq
L_{n}(\tau) & := & \frac{1}{2}\sum_{l=-\infty}^{\infty}\alpha^{\mu}_{n-l}(\tau)\alpha^{\mu}_{l}(\tau) \nn \\ 
\tilde{L}_{n}(\tau) & := & \frac{1}{2}\sum_{l=-\infty}^{\infty}\tilde{\alpha}^{\mu}_{n-l}(\tau)\tilde{\alpha}^{\mu}_{l}(\tau)
\label{eq:Lnm}
\eeq
for which we find, after repeatedly applying~(\ref{eq:anicreaNambu}) and using the Leibnizian rule
for the causal Poisson bracket, a two-time generalization
of the Virasoro algebra which explicitly reads
\beq
\hspace{-3ex}
(L_{n}(\tau),L_{m}(\tau'))=i(n-m)e^{-2in(\tau-\tau')}L_{n+m}(\tau')  \,.
\label{eq:virasorotaus}
\eeq
As expected, this algebra also reduces to the standard Virasoro 
algebra in the equal-time limit, that is, $\lim_{\tau\rightarrow\tau'}(L_{n}(\tau),L_{m}(\tau'))=\{L_{n}(\tau),L_{m}(\tau)\}=
i(n-m)L_{n+m}(\tau)$.  The generalized Virasoro algebra 
resembles in some sense the atavistic algebras studied in 
detail in 
references~\cite{FairleZachos1,FairleZachos3} but,
in our case, algebra~(\ref{eq:virasorotaus}) depends explicitly on 
two different values of the time parameter $\tau$, and not only 
on the discrete parameters $n$ and $m$, in opposition.

Finally, we note that if we write, for example,  
the causal Green's 
function for the open string~(\ref{eq:causalabierta}) in terms 
of imaginary exponentials, and by means of the expansion 
$\ln{(1+x)}=\sum_{n=1}^{\infty}((-1)^{n+1}/n)x^{n}$ for 
$\left|x\right|<1$, we may write 
\beq
& & \tilde{G}_{\mathrm{open}}(\sigma,\sigma';\tau,\tau') \nn\\
& = &  \tau-\tau' -\frac{1}{4i}\ln{\left(1-e^{2i((\tau-\tau')+(\sigma-\sigma'))}\right)}
-\frac{1}{4i}\ln{\left(1-e^{2i((\tau-\tau')-(\sigma-\sigma'))}\right)} 
\nn \\
 & & +\frac{1}{4i}\ln{\left(1-e^{2i(-(\tau-\tau')+(\sigma-\sigma'))}\right)} 
+\frac{1}{4i}\ln{\left(1-e^{-2i((\tau-\tau')+(\sigma-\sigma'))}\right)} \,,
 \label{eq:causalabierta2}
\eeq
setting the logarithmic behaviour of 
$\tilde{G}_{\mathrm{open}}(\sigma,\sigma';\tau,\tau')$ which resembles the standard Feynman propagator~\cite{gleb}, 
$G_{\mathrm{F}}(\sigma,\sigma';\tau,\tau')$, as stated at the end 
of Section~\ref{sec:CovPoissonStructure}.

\subsection{Nonlinear model}
\label{sec:Nonlinear}

As we mentioned before, we may introduce the causal 
bracket~(\ref{eq:CovBracket}) in an appropriate manner 
for the analysis of nonlinear examples.
In this section we consider the one-dimensional Lagrangian
defined by
\beq
\mathcal{L}:=\frac{1}{2}\dot{x}^2-\frac{1}{2}x^2+\frac{1}{3}
x^3 \,.
\eeq  
In Physics, this Lagrangian has been studied as 
describing  
the motion of a particle in a quasi-isochronous  storage 
ring in~\cite{ring}.  The equation of motion for this model 
reads
\beq
\dot{x}^2 = \frac{2}{3}x^3 -x^2 +2E  \,,
\eeq
where the constant term (associated to the energy of the 
particle) comes from a first integral of motion.  
Even though this equation is inherently nonlinear, we may 
introduce a well-behaved Green's function which is 
solution to the 
equation 
\beq
\frac{d^2 G(t)}{dt^2}-G^2(t)+G(t)=\delta(t)\,.
\eeq 
and explicitly given by 
\beq
\label{eq:GreenRing}
G(t)=\theta(t)\wp\left(\frac{t}{\sqrt{6}},6,-12E\right) \,,
\eeq
where $\wp$ stands for the $\wp$-Weierstrass elliptic 
function with elliptic invariants $g_2=6$ and 
$g_3=-12E$, and the function $\theta$ corresponds to the 
Heaviside step function.  In order to define our causal 
bracket we may consider the causal Green's function 
$\bar{G}(t,t')$ associated to~(\ref{eq:GreenRing}) which results
\beq
\bar{G}(t,t'):=\theta(t-t')\wp\left(\frac{t-t'}{\sqrt{6}},6,-12E\right) - 
\theta(t'-t)\wp\left(\frac{t-t'}{\sqrt{6}},6,-12E\right)  \,.
\eeq
Finally, 
the causal bracket for this model is simply given by 
\beq
(x(t),x(t'))=\bar{G}(t,t') \,.
\eeq

\section{Concluding remarks}
\label{sec:conclu}

In quantum field theory, the product of 
field operators at different spacetime points 
is well-defined.  This product, 
from the perspective of deformation 
quantization, may be extended to a star-product 
from which one defines the commutator of two quantum field operators
at different spacetime points.  This may be done
in a complete covariant way.  Thus,
taking the deformation quantization as our guiding 
programme, we have focused on the construction of a classical Poisson structure inherited from this 
quantum commutator.  To this end, we 
have considered the Green's function method 
in order to map fields to points belonging to a single
hypersurface.  
Therefore, we have found a well-defined star-product for the fields at two different spacetime 
points.  This star-product induces a classical 
causal bracket which follows the 
axioms of a Poisson structure, and may be extended trivially to obtain a bracket in the 
appropriate phase space.  
In the case of a linear system the Green's functions involved in the  bracket may be constructed explicitly, while in 
the nonlinear case the Green's function, even if we 
are able to construct it analytically,  
must be understood as associated to approximate solutions
for a given system, thus 
depending on the chosen boundary conditions and on a short spacetime parameters expansion.
Also, due to the properties of the Green's function,
in both cases the classical
bracket introduced reduces to the standard 
Poisson bracket on the assumption that our 
two spacetime points lie on the same spatial 
hypersurface, that is, in the equal-time limit.

For the case of theories involving interacting fields, we have encountered a generalization 
of Wick's theorem for the star-product of fields
at different spacetime points.  
Besides, the connection
of our formalism with standard Feynman propagator
was encountered by an appropriate isomorphism 
between star-algebras.

We have tested our formalism for typical 
models showing interesting physically motivated features. 
On the one side, we analyzed  a couple of well-known 
field theoretical models:  The 
real scalar field for which we have deduced 
a generalization of the Poisson bracket 
relations of the classical 
coefficients associated to the quantum 
creation and annihilation operators at two different spatial hypersurfaces.  
This generalization may be straightforwardly extended to the quantum counterpart. We also 
have studied the bosonic string.  In this case we
have encountered a generalization of the 
known Virasoro algebra at two different spacetime
points.  For both models, the introduced 
causal generalizations reduce to the standard 
results found in the literature at the equal-time
limit.  On the other side, we explored a nonlinear model 
for a particle in a quasi-isochronous storage ring.
For this model, we were able to find a Green's function
which allowed us to introduce the causal bracket.

Despite our results, further work has to 
be done in the direction of 
nonlinear field theories as there are several 
methods for finding Green's functions for this sort of 
theories, and it is not clear to us at this moment which
method will be more plausible to incorporate within 
our proposed bracket.  In particular, we will also need 
to understand the relevant intervals of convergence for which the 
integrals involving the Green's function
result appropriate to guarantee a well-behaved causal bracket.  
Another interesting direction will be to implement 
the causal bracket for the case of 
singular Lagrangians.
Constrained systems in the context of quantization
deformation were analyzed in~\cite{Antonsen,
Antonsen1}.  
This will be done elsewhere.

\section*{Acknowledgments}
J.B-M. 
 was supported by a CONACYT Fellowship
(Retenci\'on 206727).  J.B-M. also 
acknowledges support from PROMEP-UASLP under the 
NPTC program. AM acknowledges financial support from PROMEP
UASLP-PTC-402 and from CONACYT-Mexico under project CB-2014-243433.

\appendix

\section{Mathematical properties of the classical 
causal bracket}
\label{sec:MathProps}

Let $F_{a}:=F_{a}[\Phi(x_{k})]$ denote
the $a$-th functional on covariant 
phase space $\mathcal{S}$ attached to the
spacetime point 
$x_k\in\mathcal{M}$.
We will 
adopt the short notation 
for the classical causal Poisson 
bracket defined through deformation quantization in~(\ref{eq:CovBracket})
\begin{eqnarray}
F_{1,M}\tilde{G}^{MN}F_{2,N} & := &
\left( F_{1}[\Phi(x_{1})],F_{2}[\Phi(x_{2})]\right)   \nn\\
& = &  
\int_{\mathcal{M}} dxdx'\frac{\delta F_{1}[\Phi(x_{1})]}{\delta\Phi^{M}(x)}\tilde{G}^{MN}(x,x')\frac{\delta F_{2}[\Phi(x_{2})]}{\delta\Phi^{N}(x')}  \,,
\label{eq:ShortBracket}
\end{eqnarray}
where we have obviated the explicit dependence on the causally connected spacetime 
points 
in the manifold $\mathcal{M}$ on the left hand side of this relation.  
Here, we want to show that this bracket satisfies the axioms of a Poisson algebra as previously stated.  
First, we note that 
the bracket~(\ref{eq:ShortBracket}) is 
skew-symmetric, that is, $
\left( F_{1},F_{2}\right)=-\left( F_{2},F_{1}\right)$,
due to the skew-symmetric nature of the causal Green's function, namely, 
$\tilde{G}^{MN}=-\tilde{G}^{NM}$. 
Next, the bilinearity is obtained from the linearity of the functional derivatives involved
in~(\ref{eq:ShortBracket})
\beq
\left( F_{1},F_{2}+\alpha F_{3}\right)
& = &  F_{1,M}\tilde{G}^{MN}(F_{2}+\alpha F_{3})_{,N}
\nn\\
& = & F_{1,M}\tilde{G}^{MN}F_{2,N}
+\alpha F_{1,M}\tilde{G}^{MN}F_{3,N}  
\nn\\
& = &  \left( F_{1},F_{2}\right)+\alpha\left( F_{1},F_{3}\right) \,,
\eeq
for $\alpha$ constant.
The Leibniz rule is also directly obtained form the Leibniz rule property of the functional derivative 
\beq
\left( F_{1},F_{2}F̣_{3}\right)
& = & F_{1,M}\tilde{G}^{MN}(F_2 F_3)_{,N}
\nn\\
& = & F_{1,M}\tilde{G}^{MN} (F_{2,N}F_3+F_2 F_{3,N})
\nn\\
& = & (F_{1,M}\tilde{G}^{MN} F_{2,N})F_3
+ F_2 (F_{1,M}\tilde{G}^{MN}F_{3,N})  \nn\\
& = & \left( F_{1},F_{2}\right)F_{3}+F_{2}\left( F_{1},F_{3}\right)  \,.
\eeq
Finally, the Jacobi identity which in the adopted 
short notation reads 
\beq
P(F_{1},F_{2},F_{3})
&:=&
\left( F_{1},\left( F_{2}, F_{3}\right)\right)+\left( F_{2},\left( F_{3}, F_{1}\right)\right)+\left( F_{3},\left( F_{1}, F_{2}\right)\right)
\nn\\
&=&F_{1,LP}F_{2,M}F_{3,N}\left(\tilde{G}^{LM}\tilde{G}^{NM}+\tilde{G}^{MP}\tilde{G}^{NL} \right)
\nn\\
& & + F_{1,L}F_{2,MP}F_{3,N}\left(\tilde{G}^{MN}\tilde{G}^{LP}+\tilde{G}^{NP}\tilde{G}^{LM} \right)
\nn\\
& & + F_{1,L}F_{2,M}F_{3,NP}\left({G}^{NL}\tilde{G}^{MP}+\tilde{G}^{LP}\tilde{G}^{MN} \right), 
\eeq
may be demonstrated by using the skew-symmetry 
of the causal Green's function $\tilde{G}^{MN}$, together with the commutativity of the functional derivatives, $F_{,MN}=F_{,NM}$, for all functional $F$.  Thus, it is straightforward to show that all expressions in the last equality are vanishing.  For example, the first line in 
the last equality stands for
\beq
\nn
F_{1,LP}F_{2,M}F_{3,N}\left(\tilde{G}^{LM}\tilde{G}^{NM}+\tilde{G}^{MP}\tilde{G}^{NL} \right)  \nn\\
\hspace{10ex}=  -F_{1,PL}F_{2,M}F_{3,N}\left(\tilde{G}^{ML}\tilde{G}^{NM}+\tilde{G}^{PM}\tilde{G}^{NL} \right)  =  0\,.
\nn
\eeq
Thus, Jacobi identity also holds.  As
the four properties have been shown, we conclude 
that our causal bracket is a genuine Poisson bracket.

\end{document}